\documentclass[conference]{IEEEtran}
\IEEEoverridecommandlockouts
\usepackage{cite}
\usepackage{amsmath,amssymb,amsfonts}
\usepackage{algorithmic}
\usepackage{graphicx}
\usepackage{textcomp}
\usepackage{xcolor}
\usepackage[acronym]{glossaries}
\def\BibTeX{{\rm B\kern-.05em{\sc i\kern-.025em b}\kern-.08em
    T\kern-.1667em\lower.7ex\hbox{E}\kern-.125emX}}

\newacronym{der}{DER}{distributed energy resources}
\newacronym{ml}{ML}{machine learning}
\newacronym{dr}{DR}{demand response}
\newacronym{pv}{PV}{photovoltaics}

\begin{document}

\title{Connecting Distributed Pockets of Energy Flexibility through Federated Computations: Limitations and Possibilities\\
}

\author{\IEEEauthorblockN{Javad Mohammadi}
\IEEEauthorblockA{
%Dept. of Electrical and Computer Engineering \\
Carnegie Mellon University\\
Pittsburgh, PA, USA \\
jmohamma@andrew.cmu.edu}

\and

\IEEEauthorblockN{Jesse Thornburg}
\IEEEauthorblockA{Grid Fruit\\
Pittsburgh, PA, USA \\
jesse@gridfruit.com}
}

% \and
% \and
% \and
% \and
% \and

\maketitle

\begin{abstract}
Electric grids are traditionally operated as multi-entity systems with each entity managing a geographical region. Interest and demand for decarbonization and energy democratization %across the Western world 
is resulting in growing penetration of controllable energy resources. In turn, this process is increasing the number of grid entities. The paradigm shift is also fueled by 
%growing penetration of distributed clean energy resources as well as 
increased adoption of   
%for balancing energy supply and demand and 
%being instrumented by 
intelligent sensors and actuators equipped with advanced processing and computing capabilities. %These sensors and controllers communicate over established channels and Internet-of-Things networks and enables coordination at scale.
%The unprecedented computation and communication  
While collaboration among power grid entities (agents) reduces energy cost and increases overall reliability, achieving effective collaboration is challenging. The main challenges stem from the heterogeneity of system agents and their collected information. Furthermore, the scale of data collection is constantly increasing and many grid entities have strict privacy requirements. Another challenge is the energy industry's common practice of keeping data in silos. Federated computation is an approach well suited to addressing these issues that are increasingly important for multi-agent energy systems. Through federated computation, agents collaboratively solve learning and optimization problems while respecting each agent's privacy and overcoming barriers of cross-device and cross-organization data isolation. In this paper, we first establish the need for federated computations to achieve energy optimization goals of the future power grid. We discuss practical challenges of performing multi-agent data processing in general. Then we address challenges that arise specifically for orchestrating operation of connected distributed energy resources (DERs) in the Internet of Things (IoT). We conclude this paper by presenting a novel federated computation framework that addresses some of these issues, and we share examples of two initial field test setups in research demonstrations and commercial building applications with Grid Fruit LLC.

%As power grids become increasingly instrumented in building the smart grid, they generate data at a new scale that requires distributed optimization. As networks subject to system and node failures, power grids increasingly rely on distributed energy resources (DERs) and need \gls{ml} to manage these DER assets optimally in real time. More specifically, federated learning is an \gls{ml} technique ideal for aggregating and continuously updating distributed energy efficiency capacity across the grid.

\end{abstract}


\vspace{-.2cm}
\section{Introduction}
\vspace{-.2cm}
%\subsection{Motivation}
The future electric delivery infrastructure is expected to differ from the existing system by increased penetration of Distributed Energy Resources (DERs) for example solar photovoltaic (PV) systems, electric vehicles (EVs) and energy storage systems. DERs are multiplying across the Americas, Europe, and Africa \cite{DosSantos18}, \cite{sergi2018}. The rollout is accompanied by widespread adoption of advanced sensing, actuation, computation and communication technologies. 
While the physical infrastructure and energy exchanges will continue to be administrated and overseen by the grid operator, a central entity, the architecture of the future grid is being designed to accommodate increasing demand for energy democratization and end-use engagement \cite{ThornburgKrogh2017}. 
%there is a  for democratizing energy consumption, storage and production by end-users and third-party intermediary entities. 
Electrical utilities and regulators are updating many long-standing practices to meet these demands and interests. For example, the architecture of nascent Transactive Energy (TE) systems \cite{kok2019guest} allows a wide range of power grid entities (from homeowners with solar PV to grid operators) to collaborate and compete to provide cost-effective electricity with high reliability \cite{tushar2020peer}. 

Flexibility in production and consumption (also known as prosumption) will play a critical role in preserving reliability and resilience of the future power grids that are characterized by high penetration of intermittent energy sources (e.g., solar PV) \cite{imteaj2019leveraging}, \cite{thornburg2018}. Here, flexibility refers to the ability of a DER to adjust prosumption levels in a desired amount of time.

In parallel, resilience challenges face the future grid as cybersecurity and environmental threats are growing \cite{quirin11}, \cite{yale19}. Power system resilience is typically defined as the capability of a system to maintain its functionality and characteristics after a disturbance \cite{holling1973resilience}. In a power grid context, a disturbance often manifests as a brownout (drop in voltage) or blackout (power outage) \cite{Thornburg2015}. Identifying and aggregating flexibilities offered by power grid entities is commonly viewed as the viable solution for avoiding these disturbances and operating a reliable, resilient and climate friendly grid \cite{mohammadi2016distributed}. 

Federated computation is an approach key to both optimal DER dispatch and energy resilience in the future grid. Federated analytics and collaborative, distributed decision-making will act as the glue to connect different pieces of the future grid together, especially those connected in the Internet of Things (IoT). In other words, federated computations enable leverage IoT-connected DERs and entities (with near-ubiquitous communication and computation capabilities) to aggregate spatiotemporal flexibility of geographically-dispersed DERs. This aggregation allows control at a scale to provide this future grid with much-needed energy resilience \cite{thornburg2018}.

%\subsection{Related Works}
Federated computations in multi-agents systems are studied in different contexts. In this regard, distributed optimization and decentralized learning are established directions. The interest in these methodologies is driven by the need to develop scalable optimization and learning solutions to deal with large datasets \cite{cevher2014convex}, address privacy and security concerns \cite{yan2012distributed}, and manage multi-agent systems \cite{chen2012diffusion}.
The architecture of existing distributed optimization methods falls into two categories based whether the methods depend on a central agent to oversee computations and optimization. % \cite{kargarian2016toward}. Fully distributed methods rely on peer-to-peer information exchange. In such setups, each agent computes local updates and exchanges information with neighboring agents \cite{mohammadi2016distributed}. The communication network varies depending on the use case, and information may be exchanged in synchronous or asynchronous fashion. Consensus-based methods are among the most popular approaches for enforcing agreement among agents and directing them to achieve an optimal solution for the underlying optimization problem \cite{chang2014distributed,tsianos2012consensus}.
In the case where a central agent is responsible for regulating updates and computation among agents, all other agents are required to directly share updates with that central entity. As an example, Alternating Direction Method of Multipliers (ADMM) is a commonly used method that decomposes the original problem into sub-problems and a master problem. The ADMM setup requires each agent to solve an optimization sub-problem and share updates with the central agent. The central agent then updates the update functions of the master problem \cite{boyd2011distributed}. 

Distributed optimization and distributed learning are closely related. Distributed learning methods orchestrate cross-agent computations that enable collaboration in mining large datasets. These models often map the multi-agent setups to graph environments where nodes represents learning agents. Data may reside on nodes or edges of the representative graph \cite{mohammadi2019collaborative}. Recent works in this regards have investigated using proximal gradient methodologies \cite{li2014communication}, ADMM-based methods \cite{huang2019dp}, or stochastic gradient decent (SGD) approaches \cite{xing2015petuum}, 

%Multi-task learning methods seek to drive and share insights between related tasks to improve performance. To avoid information sharing between dissimilar tasks which decreases overall performance, researchers have proposed a wide-range of methods for assessing task similarities and knowledge sharing. Specifically, authors in 

Federated analytics has received tremendous attention recently. Federated learning is a multi-agent setup thorough which multiple entities collaborate to solve a machine learning problem. A central coordinator often regulates this process and communicates with the system's agents. Each entity keeps its raw data in confidence, performs local updates, and shares model updates only with the central entity. This iterative process allows cross-device and cross-institution model training \cite{kairouz2019advances,li2020federated}. Edge computation has been studied and implemented in limited capacities (e.g., with fog computing and query processing \cite{bonomi2012fog,deshpande2005model,garcia2015edge}). However, increasing computational and data storage capabilities of the edge devices allow inference and training tasks to be conducted locally \cite{li2020federated}. Using federated analytics across agents with heterogeneous hardware and statistical data is an active research area \cite{li2019fair}.

Organizations are often reluctant to share their data, even though sharing the knowledge and resulting insights benefits all involved parties. For example, anti-money-laundering initiatives will greatly benefit from mining the data of different financial institutions. However, requesting access to sensitive client information is an ever-present obstacle. Federated learning can enable joint financial analysis among these institutions without sharing any customer data. Another example is the collaborative efforts to reduce carbon emissions. Large building portfolios are major contributors to greenhouse gas (GHG) emissions, however management firms are not willing to share their energy and environmental data. They fear that doing so will advantage their competitors by revealing information about their building occupancies. 

Federated analytics is an effective tool for breaking these cross-device and cross-institutional silos. The approach promises to be extremely useful in the context of future energy systems with numerous IoT-connected DERs connecting and disconnecting from the grid at any given time, especially with the DERs in different locations and under different supervision and organization. 

{\color{gray}
}

In this paper, we first discuss how the electric power grid is evolving towards a multi-agent system with heterogeneous agents. From literature, including prior research works, we establish the need for federated computation across ever-increasing energy resources in section II. We begin section III by highlighting the challenges of implementing federated computation framework in IoT setups, specifically in IoT-connected DERs. We then build on \cite{mohammadi2016distributed,mohammadi2019collaborative} and presents a novel federated computation method that is well suited to address the needs of future multi-agent grid. 
%These innovations are discussed in two contexts. First, this paper introduces a distributed federated computation algorithm with centralized oversight tailored towards identifying energy flexibility of DERs and aggregating these pockets of flexibility to provide resilience for the power grid. 
The proposed method enables agents to collaboratively perform a personalized and agent-specific computation while maintaining data privacy.
%This process involves (i) leveraging oversight of a centralized agent to process centrally available information and (ii) smoothing model parameters across agents with similar specifics (i.e., similar tasks and similar data distributions).
%  The second innovation of this paper is outlining how fully decentralized learning and computations can be practically realized in a cross-device setup to achieve energy efficiency. Majority of existing federated learning methods proposed in the literature will face difficulties in real-world setups since researchers do not have access to large scale testing setups \cite{kairouz2019advances}. This motivates presenting deployment issues and accounting for them while conducting simulations. 
% Throughout the work privacy and communication bottlenecks are addressed, as these two concerns typically threaten federated learning setups \cite{kairouz2019advances}.
In section IV, we describe implementations possibilities of the above-described methods in research (through the CarnegiePLUG testbed) and commercialization (partnering with grid services company Grid Fruit LLC).

\vspace{-.15cm}
\section{The Future Plug-and-Play Grid}
\vspace{-.15cm}
The electric power grid is one of the most complex man-made systems in the world \cite{bao2009analysis}. This system spans multiple states, countries, and entire continents. Control responsibilities for this network are shared among different entities. The decision-making process occurs at different scales and across multiple spatiotemporal dimensions. The energy management paradigms of the electric grid are traditionally designed to regulate and manage a limited number of decision-making entities. Entities often control assets in a specific geographical area \cite{wood2013power}. Interactions between these entities depend on the policies and rules governing power grid operation in these areas as well as prearranged bilateral agreements. The lack of coordination to optimize cross-entity energy resources inevitably results in suboptimal system dispatch. 

% \subsection{Old grid}
% {\color{gray}
% In both cases, the task of optimally scheduling generation as well as other tasks within a control area are carried out using an Energy Management System (EMS). The coordination among neighboring control areas is generally done such that the areas agree on a flow on their tie lines and then schedule the supply of the remaining loads according to their dispatch procedure. 
% }

The future electric power grid infrastructure is emerging with a complexity that requires it be composed of many interacting elements. As the grid evolves, its new architecture should accommodate and manage an ensemble of constituent parts while ensuring interoperability and resilience. The future grid should enable plug-and-play engagement and manage heterogeneous energy resources under stress and uncertainty \cite{widergren2019grid}. The main trends behind this transition include evolving consumer expectations and technological progress. Economies of distributed electric networks are emerging at the same time, as well as growing needs for resilience given cybersecurity and environmental threats \cite{imteaj2019leveraging}.

Customers expect increased access to information and a high level of service from the twenty-first century power infrastructure. These expectations are elevated partly because outside the power sector, third party providers are offering increased choice and ease of use. For example, Uber and Lyft have disrupted the transportation system while improving it in certain respects (e.g., market-driven availability and responsiveness). On the energy side, customers demand improvements in information access as well as energy choice, equity, and reliability from their electric infrastructure. Electric grids are now expected to accommodate these rising expectations.

Technological advancements have led to ubiquitous computation and communications. This enables local information processing and decision-making and empowers customers to manage their on-site energy resources (e.g., solar PV and storage) and decide how and when to share their excess energy. Technology improvements have also unlocked new avenues for energy efficiency. For example, real-time access and coordination of building heating and cooling systems reduces their energy use and improves occupant comfort. On the other hand, the sensing and communication capabilities being added to these devices (across the spectrum of generation and consumption) creates a large attack surface for cyber intruders and can jeopardize the operations of the grid.

The energy democratization exceptions (i.e., resisting the dominant energy agenda while reclaiming and democratically restructuring energy regimes \cite{stephens2018operationalizing})  have gained steam in recent years due to access to a wider range of energy choices, widespread adoption of distributed energy resources, increased affordability of electric energy storage, and climate change considerations \cite{keshan2016}. These exceptions combined with innovative business models are encouraging local and peer-to-peer energy transactions. This shift is changing the economics of electricity generation and use.

The reliability and resilience of the aging power infrastructure is increasingly challenged by climate change, natural disasters, and cyber threats. The electric grid is key for operating other critical infrastructures.  %\cite{amini2019sustainable}. 
More specifically, any power disruption is harmful to national security, economic prosperity, and communications networks as well as the basic standard of living. The number of extreme natural weather events like wildfires, floods, storms, and heatwaves is on the rise \cite{quirin11}, \cite{yale19}. These environmental strains highlight the need for energy resiliency. For example, recent wildfires in California have pushed the state's largest utility, Pacific Gas and Electric, close to bankruptcy \cite{wsjRES}. A resilient grid is prepared to handle disruptions in the supply side caused by natural hazards, deal with cyber-physical incidents, and manage human interventions. Grid operators used to dispatch designated generation resources to preserve reliability at the time of needs. The ongoing grid transformation have enabled grid operators to access a fleet of small scale energy resources on the supply and demand sides that if orchestrated effectively can significantly boost grid resilience.
These systemic changes amplify the need for multi-entity coordination. This relies on local communication and computation capabilities. Effective coordination preserves privacy and scale across multi-modal data sources and heterogeneous devices.

\vspace{-.1cm}
\section{Scalable coordination of underutilized Energy Flexibility Pockets}
Energy flexibility is the ability of energy producers or consumers to adjust their output power (generation by producers) or their energy consumption level (demand of consumers, e.g. shifting or reducing loads) \cite{junker2018characterizing}. Energy flexibility comes in many forms and presents itself in bulk power grids as well as microgrids and urban electric systems \cite{ThornburgACM}. One example is homeowner willingness to adjust the temperature setpoint(s) of a building's cooling systems in response to a load reduction request from the grid manager. Certain strained electrical utilities send out these requests in peak energy use hours. Unless the grid takes direct control of the load in question, the energy flexibility is dependent on end-user cooperation. Another example of energy flexibility is regulating power output of a solar farm to help the grid operator stabilize the bulk power grid during extreme weather events. 

Utilizing energy flexibility offered by small scale DERs is more challenging than leveraging flexibility of large consumers or producers.
This is because DERs are constantly growing in number, in geographical spread, and in heterogeneity. If used effectively, distributed pockets of energy flexibility offered by DERs can be a reliable resource for grid management given their closer proximity to end-use facilities when compared to the grid manager (central) node. Furthermore, the distributed nature of DERs adds to the grid's flexibility for control across the network \cite{ThornburgKrogh2016}, \cite{Ustun2016}. Scalable aggregation is the key in unlocking the massive potential of DERs.

\vspace{-.15cm}
\subsection{Challenges}
\vspace{-.1cm}
IoT-connected DER controllers have a wide-range of communication and computation capabilities. This heterogeneity poses a challenge for utilizing federated computations to orchestrate DER functionalities. Put differently, data storage limitations, communication overhead, and computational constraints can degrade the performance of the analytics performed on local devices \cite{imteaj2020federated}. Moreover, the heterogeneity and multi-modality of the data collected by IoT-connected DERs pose additional difficulties in performing computations across these devices. Architectural design and computation process on edge IoT devices are extensively studied in the literature\cite{chen2019joint,chen2018federated,das2019privacy}. In what follows, we elaborate on key issues and discuss how they apply to the multi-agent DER context.

Agent hardware limitations, including constrained data storage and restricted computational and communication capabilities, can negatively affect cross-agent collaboration in multi-agent settings. These limitations slow down local computational process by increasing communication overhead and hindering intra-agent coordination. Hence, hardware specifications must be taken into account when implementing federated computations. For example, authors in \cite{hard2018federated} study hardware requirements for implementing next word prediction for a user typing. These hardware issues are even more visible in the context of nascent IoT-connected DERs.
The standards and requirements for the DER hardware setups are not yet well-defined. Manufactures and vendors are pursuing different communication protocols (e.g. OpenADR \cite{mcparland2011openadr}) while experimenting with computational possibilities. A gap also persists between access to device monitoring and measurement information in the same class of assets (e.g., solar PV inverters). 

Quality and frequency of information exchange among agents of a multi-agent network directly impact overall performance of the network. In multi-agent information processing setups, agents with limited communication resources can hinder convergence across the network. In addition, the trade-off between communication cost and frequency should be accounted for in designing federated computations \cite{ma2017distributed}. To address this issue, researchers have proposed a wide-range of solutions such as reducing the number of inter-agent information exchanges and reduce the size of exchanged message \cite{konevcny2016federated}. Accounting for communication limitations is more critical in the DERs energy management context since maintaining supply and demand balance is the key in preserving power grid's reliability. Wide-spread communication disruption can unbalance equity of generation and consumption and lead to power outages. Hence, developing robust federated computation methodologies for achieving energy management goals is critically important. 

The aforementioned issues (i.e., computation and communication constraints) leads to emergence of straggling agents. Stragglers fail to share or compute updates in time. Selective engagement of resourceful agents (computationally and communication-wise) could be a possible solution to address this issue \cite{das2019privacy}. Asynchronous implementation of updates \cite{mohammadi2015asynchronous} also alleviates this issue.
%Underutilized sources ...
Finally, depending on their location IoT-connected DERs might be governed by different policies. For example, some utilities allow DERs to provide a wide-range services for the grid while others may restrict participation to a specific aggregation arrangements. This heterogeneity of policies is an obstacle that limits coordination between DER agents located in different utility territories.

\vspace{-.2cm}
\subsection{Energy Management in DER networks}
\vspace{-.1cm}
This section intends to present a generic form for optimizing operation of a DER fleet that can be solved in a centralized fashion. The following equations formulate this problem as:

\vspace{-.1cm}
\small
\begin{subequations}\label{generalproblem}
\begin{align}
&\displaystyle \underset{{x}_{k}}{\text{minimize}}\;\; \sum_{k\in \Omega_{\text{agents}}} f_k(x_k)\label{main_obj}\\
&\text{s.t.}\hspace{1.04cm} g_j(x)\leq 0; \;\; (:\mu_j) \;\;\; j\in \Omega_{ineq} \label{ineq_cnst}\\
&\hspace{1.5cm} h_j(x)= 0;\;\;(:\lambda_j) \;\;\; j \in \Omega_{eq} \label{eq_cnst}
\end{align}
\end{subequations} 
\normalsize

In this setup the objective function ($f_k(\cdot)$) can capture a wide-range of goals including energy cost or emission minimization that $k$ agents (i.e., IoT-connected DERs) are collectively aiming to achieve. Also, sets of all agents, inequality constraints, and equality constraint are represented by $\Omega_{\text{agents}}$, $\Omega_{ineq}$, and $\Omega_{eq}$ , respectively. Moreover, $x_k$ denotes the variables associated with agent $k$. 
Inequality and equality constraints associated with each agent are denoted by  $g_j(\cdot)$ and $h_j(\cdot)$. Variables $\mu_j$ and $\lambda_j$ are Lagrangian multipliers associated with inequality and equality constraints, respectively.

Numerous solution approaches exist that are based on finding a solution that satisfies the first-order optimality conditions of this problem. % \cite{Mohammadi2016}. 
These conditions can be derived from the Lagrangian function of original optimization problem \eqref{generalproblem}. Henceforth these conditions will be denoted by $\mathfrak{O}$.

\vspace{-.2cm}
\subsection{Agent-Based Analytics}
\vspace{-.1cm}
This section presents a novel federated computation framework which directly solves the first order optimality conditions of the underlying optimization problem (i.e., \eqref{generalproblem}). Thus, technically reduces the optimization problem to finding a solution for a coupled system of equations with geometric constraints. This system of equations can be solved in a fully distributed manner through an iterative process. This is due to the fact each equation only contains local variables (i.e., related an agent or its neighboring agents)

This iterative process relies on the agents to conduct local updates and information exchange. In the case that a problem formulation includes constraints of the physical power network, physical-connections needs to be reflected in the communication graph. Each agent $i$ is only required to update variables associated with itself (namely $V_{i}$) and share updates with a limited number of other agents. Collectively, agents of this multi-agent system solve the optimality conditions of the underlying optimization problem. In order to solve the discussed set of equation, the federated computation framework creates local copies of shared variables and assign each copy to a corresponding agent \cite{olfati2007consensus}. %,olfati2005consensus}. 
The iterative process ensures that all the local copies converge to the same value. The formula for local updates of each agent is presented below, where the consensus term enforces agreement between local copies of agents. The final term uses the updated copies to solve the optimality conditions.

\vspace{-.2cm}
\small
 \begin{eqnarray}
V_{i}(k+1)&=&\mathbb{P}[V_i(k)+{\boldmath\rho}_i^\mathcal{C}(k) \overbrace{\sum_{j\in\Omega_i}(V_i(k)-V_j(k))}^{\text{neighborhood consensus}}\label{y_update}\\  \nonumber && +{\boldmath\rho}_i^\mathcal{I}(k) \underbrace{\mathfrak{O}_i(V_i(k))}_{\text{optimality conditions}}]_\mathcal{F}, \; \; j\in \Omega_i
  \end{eqnarray}
\normalsize

% \begin{equation}
% V_{i}(k+1)=\mathbb{P}[V_i(k)+{\boldmath\rho}_i^C \sum_{j\in\Omega_i}(V_i(k)-V_j(k))+ 
% {\boldmath\rho}_i^I \mathfrak{O}(V_j(k))]_\mathcal{F}, \; \; j\in \Omega_i
%   \label{y_update}
%   \end{equation}

In \eqref{y_update}, $V_{i}$ is the collection of variables associate with agent $i$ including $x_i,\mu_j, \lambda_j$. Also, $k$ denotes the iteration number whereas $\Omega_i$ represents the neighboring set of agent $i$. 
Furthermore, $\mathfrak{O}_i(\cdot)$ represents the first order optimality constraints associated with agent $i$ while $\rho_i^C$ and $\rho_i^I$ denote the tuning parameter vectors. Finally, $\mathbb{P}$ is the projection operator that projects $V_i$ onto its determined feasible space, denoted by $\mathcal{F}$. The feasible space is defined as the intersection of equality and inequality constraints of \eqref{generalproblem}.

Updating variables in \eqref{y_update} only involve algebraic functions. The lightweight linear structure of these federated updates is well-suited to work on resource-constrained DER controllers. This update scheme can also be implemented in a fully distributed peer-to-peer fashion or distributed with centralized oversight (i.e., where a central agent receives information update from all agents).
%The update structure accommodates both synchronous and asynchronous updates which can substantially reduce the communication overhead. Finally, the proposed method accommodates a wide-range of communication architectures. 
The updates for variables of all agents follows the following compact form:

%Consequently, the update rules for the all variables at the intra-network optimization of network $\mathcal{N}$  in a dense form is provided by (\ref{intradense})

\vspace{-.15cm}
\small
\begin{align}
\mathcal{V}(k+1)&=\widetilde{\mathcal{V}}(k)-A\widetilde{\mathcal{V}}(k)+C\nonumber\\
\widetilde{\mathcal{V}}(k+1)&=\mathbb{P} \left[ \mathcal{V}(k+1) \right]_\mathcal{F}\label{intradense}
\end{align}
\normalsize

\noindent $\mathcal{V}$ contains stacked variables for all agents, where $A$ and $C$ represent system parameters and tuning weights. Moreover, $\mathbb{P}$ is the projection operator that enforces variables to stay in the feasible space (i.e., $\mathcal{F}$). Also, $\widetilde{\mathcal{V}}$ is the vector of the stacked projected variables. Convergence analysis is presented in \cite{mohammadi2016distributed}.%the Appendix.

\begin{figure}
\includegraphics[width=0.45\textwidth]{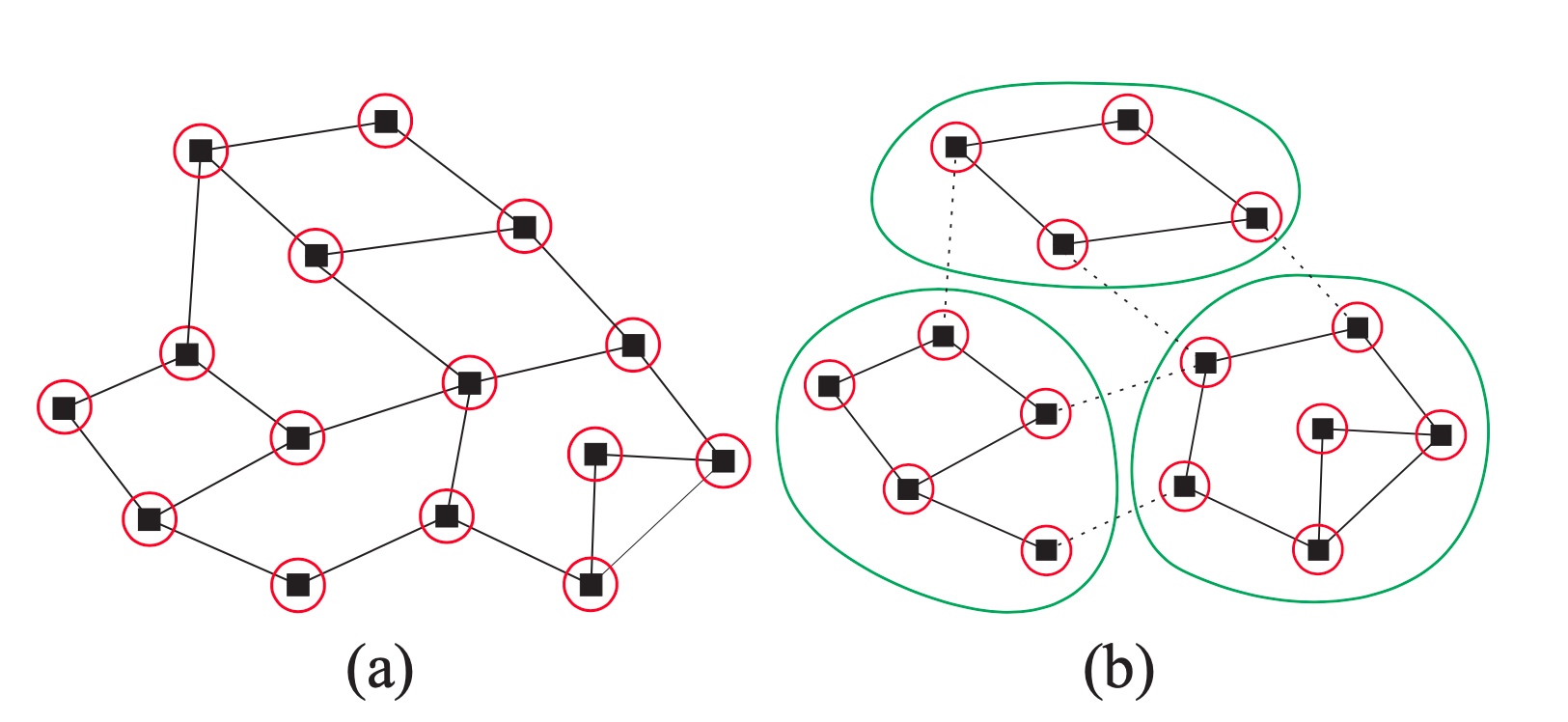}
\caption{(a) Synchronous update and (b) Asynchronous update}
\vspace{-.35cm}
\label{fig:sync_async}
\centering
\end{figure}

On a separate note, the federated computation framework being presented accommodates both synchronous and asynchronous update architectures. The difference between the two schemes is the frequency of information exchange between agents. In the synchronous update structure, every agent exchanges information with all neighboring agents after each iteration. Agents wait for each neighboring agens to send them updated information before performing the next update of their local variables.  Under the asynchronous update regime, some agents only exchange information after intervals of a few iterations. In this setup, agents are grouped into clusters where the exchange between agents within a cluster may take place every iteration while communication with agents in another cluster typically occurs every few iterations. These two update schemes are shown in Fig. \ref{fig:sync_async}.
Asynchronous architectures can substantially reduce the communication overhead \cite{mohammadi2015asynchronous} as inter- and intra-cluster updates follow two different clocks. 

\vspace{-.2cm}
 \section{Benchmark and Implementation}
 \vspace{-.1cm}
 The majority of existing federated analytics methods proposed in the literature will face difficulties in real-world setups since researchers do not have access to large scale testing setups \cite{kairouz2019advances}. This motivates developing test-bed for evaluating federated computation methods. 
In what follow, we discuss two test setup for testing federated computations in the context of future electric grids; \textit{CarnegiePLUG} and Grid Fruit, LLC.

The \textit{CarnegiePLUG} testbed at Carnegie Mellon University (CMU) is a campuswide hardware-in-the-loop (HIL) setup. \textit{CarnegiePLUG} (short for "prosumer-in-the-loop simulation grid") enables large-scale sensing, computation and actuation over a network of heterogeneous IoT-connected energy producers and consumers (prosuming) assets \cite{kim2020carnegieplug}. This system is build on a multi-agent architecture that facilitates plug-and-play integration of virtual (simulated) and on-campus DERs. \textit{CarnegiePLUG} has three layers; agent integration, communication and data acquisition. Most CMU buildings are equipped with sensors to collect fine-grained energy and environmental data (e.g.,  electricity  consumption  and temperature of campus offices and labs). \textit{CarnegiePLUG} utilizes this data in real-time and enables connecting this data to related agents in the simulation environment. In \textit{CarnegiePLUG} setup each building or room can be modeled as an agent. The communication between agents will take place in the open-source, DOE-funded
%and Department-of-Energy-supported 
$\text{VOLTTRON}^{\text{TM}}$ platform \cite{volttronLINK}.

Grid Fruit is a startup company that puts unused energy and operational data to work for the grid and its commercial buildings. Specifically, in collaboration with us Grid Fruit has developed machine learning software to provide energy efficiency and grid incentives to their commercial consumers. Using machine learning with enhanced monitoring and dynamic control, Grid Fruit optimizes load operations across each building to minimize wasted energy and emergency maintenance procedures on key building loads. In their food store projects specifically, Grid Fruit is enabling lower energy bills, operational costs, and food waste while transforming commercial refrigeration systems into thermal batteries for the grid. This, in turn, allows food stores to access demand response rebates from the grid and reduce the stores’ peak demand charges, which constitute the majority of their electricity bills in many regions.

Grid Fruit is optimizing the schedule of time-shiftable load events (e.g., defrost cycles in refrigeration systems) in multiple load types for commercial buildings. The approach described in section III can be tailored to allow Grid Fruit to optimally balance DERs including rooftop PV and multiple load types (e.g., HVAC, lighting, refrigeration). In Grid Fruit's recent project with utility customer Southern California Edison (SCE), Grid Fruit is optimizing load scheduling in convenience stores across the utility (around 5,345 stores) and across the state of California (11,990 stores) \cite{lenard20}. Federated computation aggregates load flexibility and smooths the aggregate demand of these convenience stores. The proposed project is generating and modeling load scheduling commands over year-long simulations with SCE grid conditions (e.g., SCE time-of-use pricing and demand charges). Furthermore, the federated computation approach enables optimization of stores aggregated across all of California's sixteen distinct climate zones. A key metric of interest to SCE is reduction in max demand as compared with the pre-intervention baseline. In initial tests, Grid Fruit has demonstrated reduction in maximum demand of up to 58.2\% in evening peak hours for a single store (36.6\% reduction over the entire day). 
Grid Fruit intends to apply the above-described federated learning approach to optimize operation of the convenience store loads (real-world DERs) across the entire SCE service territory and across the state of California. Aggregating the flexibility potential of food stores across these areas will turn the untapped resource of commercial building demand into a grid asset that enables peak shaving on the order of 100 MW at peak-demand hours.
%Grid Fruit has shown max demand reduction of X\% in evening peak hours (Y\% reduction over the entire day).

% \section{Preliminary results}
% \subsection{edge intelligence}
% Require smart control solutions

% Grid Fruit offer digital personalization of cooling. Cooling was not designed for comfort \cite{smil2020year} 

% \subsection{high-level formulation}
% {\color{red}Abstract Soheil's formulation}
% \subsection{Example Results}
% {\color{red}Results from SCE pilot}
% Based on pilot implementation result with Southern California Edison:
% w.o individual control -- may end up w overlapped defrost cycles
% w collaborative control -- Staggered control

% \section*{Acknowledgment}
% Authors would like to acknowledge Carolyn Goodman contributions and input and appreciate Dr. Soheil Kolouri's feedback and insights. Authors are also thankful for Electric Power Research Institute (EPRI) and Southern California Edison (SCE) for their guidance and partnership.  

\vspace{-.15cm}
\section{Conclusion}
\vspace{-.1cm}
In this paper, we have highlighted the impact and role of IoT-connected DERs in the future of the electric power grid. The number of affordable energy resources (e.g., solar PV) is on the rise. These systems come in a wide range of sizes with different monitoring and control technologies and capabilities. This inherent heterogeneity makes decision-making and coordination across these various systems challenging. Federated computation enables large-scale machine learning and optimization across heterogeneous agents. This approach is well suited to orchestrating operation of IoT-connected DERs across many locations. We have detailed the hurdles for achieving effective coordination across IoT networks, such as complications of DER controllers. We presented a novel, lightweight, coordination method and discussed implementation and benchmarking directions in both research and commercial contexts.

\vspace{-.35cm}
\bibliographystyle{./bibliography/IEEEtran}
\bibliography{./bibliography/IEEEabrv,./bibliography/IEEEexample, ./bibliography/refs}

% \appendix
% \input{appendix}

\end{document}